\documentclass[12pt]{article}

\renewcommand{\thefootnote}{\fnsymbol{footnote}}
\usepackage{amsbsy,amssymb,latexsym,amsfonts,amsmath}
\usepackage{mathrsfs}
\usepackage{graphicx}
\usepackage{bm}
\usepackage{here}
\newcommand{\bel}[1]{\begin{equation}\label{#1}}                     
\newcommand{\bal}[1]{\begin{eqnarray}\label{#1}}                     
\newcommand{\be}{\begin{equation}}
\newcommand{\ee}{\end{equation}}
\newcommand{\im}{\mathrm{i}}
\newcommand{\ex}{\mathrm{e}}
\newcommand{\de}{\mathrm{d}}

\newcommand{\qq}{\qquad}

\begin{document}
%%%%%%%%%%%%%%%%%%%%%%%%%%%%%%%%%%%%%%%%%%%%%%%%%%%%%%%%%%%%%%%%%%%%%%%%%%%%%%%%
%%%%%%%%%%%%%%%%%%%%%%%%%%%%%%%%%%%%%%%%%%%%%%%%%%%%%%%%%%%%%%%%%%%%%%%%%%%%%%%%%%%%%%%%%%
%
% title page
%
%%%%%%%%%%%%%%%%%%%%%%%%%%%%%%%%%%%%%%%%%%%%%%%%%%%%%%%%%%%%%%%%%%%%%%%%%%%%%%%%%%%%%%%%%
\begin{titlepage}
%%%%%%%%%%%%%%%%%%%%%%%%%%%%%%
\begin{flushright}
\normalsize
%\filename
~~~~
OCU-PHYS 479\\
September, 2018 \\
\end{flushright}
%%%%%%%%%%%%%%%%%%%%%%%%%%%%%%%%%%%%%%%%%%%%%%%%%%

\vspace{15pt}

%%%%%%%%%%%%%%%%%%%% title %%%%%%%%%%%%%%%%%%%%%%
\begin{center}
{\LARGE  Discrete Painlev\'{e} system and the double scaling limit } \\
\vspace{10pt}
{\LARGE of the matrix model for irregular conformal block }\\
\vspace{10pt}
{\LARGE and gauge theory }
\end{center}
%%%%%%%%%%%%%%%%%%%%%%%%%%%%%%%%%%%%%%%%%%%%%%%%%%

\vspace{23pt}

%%%%%%%%%%%%%%%%%%% authors %%%%%%%%%%%%%%%%%%%%%%
\begin{center}
{ H. Itoyama$^{a, b}$\footnote{e-mail: itoyama@sci.osaka-cu.ac.jp},
T. Oota$^b$\footnote{e-mail: toota@sci.osaka-cu.ac.jp}
  and Katsuya Yano$^a$\footnote{e-mail: yanok@sci.osaka-cu.ac.jp}   }\\

%%%%%%%%%%%%%%%%%%%%%%%%%%%%%%%%%%%%%%%%%%%%%%%%%%
%
\vspace{18pt}
%
%%%%%%%%%%%%%%%%%%% affiliation %%%%%%%%%%%%%%%%%%%

$^a$\it Department of Mathematics and Physics, Graduate School of Science\\
Osaka City University\\
\vspace{5pt}

$^b$\it Osaka City University Advanced Mathematical Institute (OCAMI)

\vspace{5pt}

3-3-138, Sugimoto, Sumiyoshi-ku, Osaka, 558-8585, Japan \\

\end{center}
%%%%%%%%%%%%%%%%%%%%%%%%%%%%%%%%%%%%%%%%%%%%%%%%%%%
%
\vspace{20pt}
\begin{center}
Abstract\\
\end{center}
%%%%%%%%%%%%%%%%%%%% abstract %%%%%%%%%%%%%%%%%%%%%
  We study the partition function of the matrix model of finite size that realizes the irregular conformal block 
 for the case of the ${\cal N}=2$ supersymmetric $SU(2)$ gauge theory with $N_f =2$.
 This model has been obtained in [arXiv:1008.1861 [hep-th]] as the massive scaling limit of the $\beta$-deformed matrix model 
 representing the conformal block. We point out that 
 the model for the case of $\beta =1$ can be recast into a unitary matrix model with log potential and show that it is exhibited as a discrete 
 Painlev\'{e} system by the method of orthogonal polynomials. We derive the Painlev\'{e} II equation, taking the double scaling limit in the vicinity of 
 the critical point
 which is the Argyres-Douglas type point of the corresponding spectral curve.
 By the $0$d-$4$d dictionary, we obtain the time variable and the parameter of the double scaled theory respectively from the sum and the difference of the two mass parameters scaled to their critical values.
 
%%%%%%%%%%%%%%%%%%%%%%%%%%%%%%%%%%%%%%%%%%%%%%%%%%%

\vfill

\end{titlepage}

%%%%%%%%%%%%%%%%%%%%
\renewcommand{\thefootnote}{\arabic{footnote}}
\setcounter{footnote}{0}
%%%%%%%%%%%%%%%%%%%%

%%%%%%%%%%%%%%%%%%%%%%%%%%%%%%%%%%%%%%%%%%%%%%%%%%%%%%%%%%%%%%%%%%%%%%%%%%%%%%%%
%%%%%%%%%%%%%%%%%%%%%%%%%%%%%%%%%%%%%%%%%%%%%%%%%%%%%%%%%%%%%%%%%%%%%%%%%%%%%%%%

  Study of correlation functions in lower dimensional quantum field theory and statistical system
 has sometimes led us to a surprising occurrence of nonlinear differential/difference equations that they obey.
 In two dimensional physical systems, these equations are typically Painlev\'{e} equations that have
 attracted interest of both physicists and mathematicians, and that govern the scaling behavior of the systems. 
  They first appeared in the study of Ising two point correlation functions, and  take the form of Painlev\'{e} III \cite{BMW73,TM73,WMTB,McW,Mc90}.
  
  The second such development was made in the context of two dimensional quantum gravity (2d gravity for short)
  and $c < 1$ non-critical strings \cite {BKDSGM}. For a review, see, for example, \cite{LAG}. 
  Equilateral triangulation of a two dimensional random surface generates the double infinite sum for its partition function 
  by the number of triangles and by the number of
   holes of the discretized surface. Through dual Feynman diagrams, the partition function is recast
  into multiple integrals of a hermitean matrix of finite size $N$ and  hence is the finite $N$ 
  hermitean matrix  model having a bare cosmological constant parameter.  The method of orthogonal polynomials
  permits us to relate the partition function with a set of recursion relations.  A specific set of recursion relations
  forms a system of difference equations called string equations, and a phenomenon of genus enhancement/resummation takes place
  in a limit commonly referred to as the double scaling limit where a difference operator is replaced by the corresponding derivative.
   Painlev\'{e} I equation has been exhibited this way as a universal equation governing the nonperturbative scaling function 
    for the partition function containing all genus contributions. 
     
     In recent years, a certain class of $\beta$-deformed ensembles for matrix models containing log potentials have been serving
     as integral representations \cite{DF,DV,IMO,EM,MMS10,IO5,IOYone} of 2d conformal and irregular 2d conformal block \cite{GaiGT,Irreg}.
     They in fact generate directly \cite{IO5,IOYone} the expansion of the block 
     in the form of the instanton expansion
       in accordance with the AGT correspondence \cite{AGT}. The matrix model free energy $F$ is thus equal to the 
       instanton part of the Seiberg-Witten prepotential ${\cal F}$ augmented by the higher genus contributions \cite{Nek}: $F = {\cal F}$.
       For a review, see, for example \cite{IY}.
      In \cite{GIL12}, Painlev\'{e} VI equation has been derived for a  Fourier transform of the $c=1$ conformal block with
      respect to the intermediate momentum.  See \cite{GIL13,ILT,Nag,BLMST,MM17,GG}  for subsequent analyses.

  In this letter,  we will study the simplest prototypical case of the irregular block, namely the case of
 the ${\cal N}=2$ supersymmetric $SU(2)$ gauge theory with the number of hypermultiplets $N_f =2$ in the form of the matrix model integral representation
  derived in \cite{IOYone}. Unlike \cite{GIL12}, our procedure is closer in spirit to that of 2d gravity
   in its unitary counterpart. We see that the finite $N$ system formulated by the orthogonal polynomials which we devise
    is already regarded as
    a discretized  Painlev\'{e} system. We are able to take the double scaling limit of this system to its critical point 
     to derive the Painlev\'{e}  II equation  for the scaling function.
     The ``time" variable $t$ is obtained from the limit of the sum of
     the two hypermultiplet masses of the gauge theory to its critical value by the $0$d-$4$d dictionary while  the parameter $M$ in the
      equation  from the limit of the difference of the two masses.  
      Details of the derivation to our findings will be given elsewhere.

%%%%%%%%%%%%%%%%%%

The partition functions of the
$\beta$-deformed matrix models which directly generate \cite{IO5,IOYone} the instanton expansion of the four-dimensional $\mathcal{N}=2$
$SU(2)$ 
gauge theories with $N_f$ fundamental matters can be generically presented as
\be
Z^{(N_f)} = \mathcal{N}_{(N_f)} 
\left( \prod_{I=1}^N \int_{\mathcal{C}_I^{(N_f)}} \de w_I \right)
\Delta(w)^{2\beta} \exp\left( \sqrt{\beta} \sum_{I=1}^N W^{(N_f)}(w_I) \right).
\ee
Here, $\mathcal{N}_{(N_f)}$ is a normalization factor,
$\Delta(w) = \prod_{I<J} (w_I - w_J)$ the Vandermonde determinant,
and $\mathcal{C}_I^{(N_f)}$ certain integration contours below.

For $N_f=4$, namely, the case of $2$d conformal block, the potential $W^{(4)}(w)$ 
is given by the three-Penner (logarithmic)  potential
\be
W^{(4)}(w) = \alpha_1 \log(w) + \alpha_2 \log(w-q_0) + \alpha_3 \log(w - 1).
\ee
Let $N=N_L + N_R$. The $N_L$ contours $\mathcal{C}_I^{(4)}$ ($1 \leq I \leq N_L$)
are taken to be the interval $[0,q_0]$ and the remaining $N_R$ contours $\mathcal{C}^{(4)}_J$
($N_L+1 \leq J \leq N$)
are chosen as
$[1, \infty]$.
This corresponds to the four-point conformal block of the two-dimensional conformal field theory with $c=1-6\, Q_E^2$, 
$Q_E \equiv \sqrt{\beta}-1/\sqrt{\beta}$:
\be
Z^{(4)} = \langle V_{\alpha_1}(0) V_{\alpha_2}(q_0) V_{\alpha_3}(1) V_{\alpha_4}(\infty) \rangle.
\ee
The $\beta$-deformed matrix model for $N_f=4$ contains seven parameters
$\alpha_1$, $\alpha_2$, $\alpha_3$, $\alpha_4$, $\beta$, $N_L$, $N_R$
undergoing one constraint (the momentum conservation)
\begin{align}
\alpha_1 + \alpha_2 + \alpha_3 + \alpha_4 + 2 \sqrt{\beta}
N  =2\, Q_E. \label{constrain}
\end{align}
These are transcribed into six unconstrained $4$d parameters of $N_f = 4$ $SU(2)$ gauge theory
\be
\frac{\epsilon_1}{g_s},~ \frac{a}{g_s},~ \frac{m_1}{g_s},~ \frac{m_2}{g_s},~ \frac{m_3}{g_s},~ \frac{m_4}{g_s},
\ee
by the $0$d-$4$d dictionary \cite{IO5}\footnote{Here in comparison to \cite{IO5}, we have renamed the mass parameters as $m_1^{[13]} = m_4$, $m_3^{[13]} = m_1$, $m_4^{[13]} = m_3$, such that
the ordering of masses subsequently sent to infinity is natural.}:
\begin{align}
\sqrt{\beta} N_L &= \frac{a - m_2}{g_s} ,& 
\sqrt{\beta} N_R &= -\frac{a + m_1}{g_s}, \nonumber\\
\alpha_1 &= \frac{1}{g_s} \left( m_2 - m_4 + \epsilon \right) , & 
\alpha_2 &= \frac{1}{g_s} \left( m_2 + m_4 \right), \label{dict} \\
\alpha_3 &= \frac{1}{g_s} \left( m_1 + m_3 \right), & 
\alpha_4 &= \frac{1}{g_s}\left( m_1 - m_3 + \epsilon \right). \nonumber
\end{align}
The omega background parameters $\epsilon_{1,2}$ are related to $\beta$ as $\epsilon_1 = \sqrt{\beta} \, g_s$
and $\epsilon_2 = - g_s/\sqrt{\beta}$. Hence $g_s^2 = - \epsilon_1 \, \epsilon_2$ and $\epsilon \equiv
\epsilon_1 + \epsilon_2 = Q_E \, g_s$.
The cross ratio $q_0$ is identified with the exponentiated ultraviolet gauge coupling constant
$q_0 \equiv e^{\im \pi \tau_0}$, 
$\tau_0 \equiv (\theta_0/\pi) + 8\,  \pi \, \im/g_0^2$.

The $N_f=3$ limit of the gauge theory is taken by $m_4 \rightarrow \infty$ with 
 $\Lambda_3 \equiv 4\, q_0 \, m_4$ fixed. 
 By the above dictionary \eqref{dict}, this corresponds to
the $q_0 \rightarrow 0$ limit
with $2 \, q_{03} \equiv q_0 \left( -\alpha_1 + \alpha_2 \right)$ and
$\alpha_{1+2}\equiv \alpha_1+\alpha_2$ fixed. 
The parameter $q_{03}$
is related to the dynamical mass scale $\Lambda_3$ of the $N_f=3$ theory by
 $q_{03}
= \Lambda_3/(4\, g_s)$. Also, $\alpha_{1+2}=(2\, m_2+\epsilon)/g_s$.
The constraint \eqref{constrain} reduces to
\begin{align}
\alpha_{1+2} + \alpha_3 + \alpha_4 + 2 \sqrt{\beta} N  = 2 Q_E . \label{constNf3}
\end{align}
In the $N_f=3$ limit, the potential $W^{(3)}(w)$ for the ``irregular matrix model''
consists of two logarithmic terms and an inverse power term:
\begin{align}
W^{(3)}(w) = \alpha_{1+2} \log w + \alpha_3 \log(w-1) - \frac{q_{03}}{w}.
\end{align}
For an explicit form of integration contours $\mathcal{C}^{(3)}_I$, see \cite{IOYone}.

We can take the $N_f=2$ limit in the gauge theory subsequently after the $N_f = 3$ limit by $m_3 \rightarrow \infty$ with the dynamical scale $\Lambda_2 \equiv (m_3\, \Lambda_3)^{1/2}$ fixed. In the matrix model, this corresponds to the $q_{03} \rightarrow 0$ limit
with $q_{02}{}^2 \equiv (1/2)q_{03} \left( \alpha_3 - \alpha_4 \right)$
and $\alpha_{3+4} \equiv \alpha_3 + \alpha_4$ fixed. The momentum conservation \eqref{constNf3}
becomes
\be
\alpha_{1+2} + \alpha_{3+4} + 2 \sqrt{\beta} N  = 2\,  Q_E. \label{constNf2}
\ee
Now $\alpha_{3+4} = (2 m_1 + \epsilon)/g_s$ and $q_{02}=\Lambda_2/(2\, g_s)$.
The potential of the resultant $N_f=2$ irregular matrix model takes the following form:
\be
W^{(2)}(w) =\alpha_{3+4} \log w 
 -q_{02} \left( w + \frac{1}{w}\right).
\label{Nf2pot}
\ee

%%%%%%%%%%%%%%%%%%%%%%%%%%%%%%%%%%%%%%%%%%%%%%%%%%

In dealing with matrix model partition functions in general, one needs to predecide whether filling fractions
are explicitly specified or not as the number of integrations lying in different contours.
It is argued in \cite{MM17} (also implicit in \cite{GIL12}) that these two distinct cases are related
to each other by a version of Fourier transform.
We will transplant their discussion at $N_f=4, 0$ here at $N_f=2$.

In general, let
\be
\underline{Z}{}_s(N) = \sum_n s^n Z(N-n,n).
\ee
Here $Z(N-n,n)$ is the object which we have been discussing up to now and $\underline{Z}{}_s(N)$
is the object which we will study from now on for the derivation of Painlev\'{e} system,
the point of view of which is in accordance with the current literature.
For simplicity, we set $s=1$ and we have one less parameters at hand from now on.

Moreover, it has been argued \cite{DDJT,DSS,CM91} that two-cut hermitean matrix model and the unitary matrix model share the critical properties.
Therefore,
the planar scaling or the double scaling limit
of our irregular models would lie in the same universality class which the unitary matrix model belongs to.

From now on, we restrict ourselves to the case $\beta=1$. Hence $\epsilon_1 = - \epsilon_2=g_s$ and $\epsilon=0$.
Let us consider the following unitary matrix model
\bel{PartZN}
\begin{split}
\underline{Z}{}_{U(N)}&=\frac{(-1)^{(1/2)N(N-1)}}{N!} \left( \prod_{I=1}^{N} \oint \frac{\de w_I}{2\pi \im} \right)
\Delta(w)^2 \, \exp\left( \sum_{I=1}^N W(w_I) \right) \cr
&=  \frac{1}{N!}\left( \prod_{I=1}^N \oint \frac{\de w_I}{2\pi \im\, w_I} \right)
\Delta(w) \Delta(w^{-1}) 
\exp\left( \sum_{I=1}^N W_{U}(w_I) \right),
\end{split} 
\ee
where the potential is given by
$W_U(w) = W(w) + N \log w$.
In particular, we study the $N_f=2$ case for simplicity:
\be
W_{U}(w) = W^{(2)}(w)+N \log w= - q_{02} \left( w + \frac{1}{w} \right)
+ M \log w, \label{potNf2}
\ee
where $M \equiv  \alpha_{3+4}+N=(m_1-m_2)/g_s$. 
In order for the contour integrals to be well-defined, we assume that $M$ is an integer.
In \cite{IOYone}, the original integration path on the real axis of the $N_{f}=4$ model was deformed into a contour in
the complex plane by an analytic continuation to avoid the singularity  which is induced by 
the $N_f =3,\, 2$ potential in the limit. In $N_f =2$ case, the contour derived is the one wrapping the positive real axis from the origin to infinity. When $M$ is an integer, this contour becomes a closed circle around the origin. Our $N_f=2$ model is
in fact equivalent to the above unitary matrix model. 

The unitary matrix model can be solved  \cite{MP90,PS90a,PS90b} by the method of orthogonal polynomials \cite{bes79,IZ80}. 
Let us use the monic orthogonal polynomials \cite{PS90a,PS90b}\footnote{In \cite{MP90},
orthogonal polynomials of different type have been introduced
to solve the unitary matrix model.}:
\be
\oint \frac{\de w}{2\pi \im \, w} \ex^{W_U(w)} \, 
 p_n(w) \, \tilde{p}_m(w^{-1}) = h_n \delta_{n,m}, 
\ee
\be
p_n(w) = w^n + \sum_{k=0}^{n-1} A^{(n)}_k\, w^k, \qq
\tilde{p}_n(w^{-1}) = w^{-n} + \sum_{k=0}^{n-1} B^{(n)}_k \, w^{-k}.
\ee
In \cite{PS90a,PS90b}, $M=0$ case was considered with $\tilde{p}_n(w)=p_n(w)$ ($B^{(n)}_k=A^{(n)}_k$).

Through explicit computation, we have found that the moments of this model are given by the modified Bessel functions up to phase factors.
Let
\be
K^{(n)}_{\nu} \equiv \det\Bigl( I_{j-i+\nu}(1/\underline{g}{}_s)  \Bigr)_{1 \leq i,j \leq n}, \qq
(\nu \in \mathbb{C};  n=0,1,2,\dotsm), 
\ee
where $I_{\nu}(z)$ is the modified Bessel function of the first kind and
$\underline{g}{}_s \equiv 1/(2\, q_{02})=g_s/\Lambda_2$. The normalization constants of the orthogonal polynomials
are given by $h_n=(-1)^{M} K^{(n+1)}_M/K^{(n)}_M$. In particular,  $h_0=(-1)^M \, I_{M}(1/\underline{g}{}_s)$.  
For notational simplicity, let $A_n \equiv p_n(0)=A^{(n)}_0$, $B_n \equiv \tilde{p}_n(0)=B^{(n)}_0$. 
These constants are respectively given by 
$A_n = K^{(n)}_{M+1}/K^{(n)}_M$,
$B_n = K^{(n)}_{M-1}/K^{(n)}_M$.

The partition function \eqref{PartZN} can be written in terms of these objects:
\bel{PartZN2}
\underline{Z}{}_{U(N)} = (-1)^{MN} K^{(N)}_M=
\prod_{k=0}^{N-1} h_k = h_0^N \prod_{j=1}^{N-1} \bigl(1 - A_j B_j \bigr)^{N-j}.
\ee

The string equations lead to
the following recursion relations for  $A_n$ and $B_n$:
\bel{ABrec}
A_{n+1} = - A_{n-1} + \frac{2\, n\, \underline{g}{}_s\, A_n}{1 - A_n\, B_n}, \qq
B_{n+1} = - B_{n-1} + \frac{2\, n\, \underline{g}{}_s\, B_n}{1- A_n\, B_n}, 
\ee
\bel{ABrec2}
A_n B_{n+1} - A_{n+1} B_n = 2\, M\, \underline{g}{}_s.
\ee
With the initial conditions $A_0=B_0=1$, and
\be
A_1 = \frac{I_{M+1}(1/\underline{g}{}_s)}{I_{M}(1/\underline{g}{}_s)}, \qq
B_1 = \frac{I_{M-1}(1/\underline{g}{}_s)}{I_{M}(1/\underline{g}{}_s)},
\ee
the remaining constants $A_n$ and $B_n$ are completely characterized by the recursion relations \eqref{ABrec}, \eqref{ABrec2}. When $M=0$ with $B_n=A_n$, the recursion relation \eqref{ABrec}
is the discrete Painlev\'{e} II equation \cite{PS90a}. Moreover, let $x_n=A_{n+1}/A_n$ and $y_n=B_{n+1}/B_n$.
Then from the recursion relations \eqref{ABrec} and \eqref{ABrec2}, one can show that these variables
respectively 
satisfy the alternate discrete Painlev\'{e} II equations \cite{FGR93,NSKGR96}:
\be
\frac{2(n+1) \underline{g}{}_s}{1+x_n x_{n+1}} + \frac{2n\, \underline{g}{}_s}{1+x_n x_{n-1}}
= - x_n + \frac{1}{x_n} + 2 \, n \underline{g}{}_s - 2 M \underline{g}{}_s,
\ee
\be
\frac{2(n+1) \underline{g}{}_s}{1+y_n y_{n+1}} + \frac{2n\, \underline{g}{}_s}{1+y_n y_{n-1}}
= - y_n + \frac{1}{y_n} + 2 \, n \underline{g}{}_s + 2 M \underline{g}{}_s.
\ee
Note that our solutions 
\be
x_n = \frac{K^{(n+1)}_{M+1} K^{(n)}_M}{K^{(n+1)}_M K^{(n)}_{M+1}}, \qq
y_n = \frac{K^{(n+1)}_{M-1} K^{(n)}_M}{K^{(n+1)}_M K^{(n)}_{M-1}}
\ee
belong to a class of the Casorati determinant solutions to the alt-dPII considered in \cite{NSKGR96}.
Furthermore, the partition function  $\underline{Z}{}_{U(N)}= (-1)^{MN} K^{(N)}_M$
is the $\tau$-function of the alt-dPII equation. It is well known that the alt-dPII equation is closely
related to the (differential) Painlev\'{e} III equation (PIII${}_1$ or PIII($D_6^{(1)}$)).
Following \cite[(4.26)]{FW02}, let us introduce a function of $t=1/\underline{g}{}_s^2$ by
\be
\sigma(t) := -t \frac{\de}{\de t} \log \Bigl( \ex^{-t/4} t^{M^2/4} K^{(N)}_M \Bigr).
\ee 
Then, $\sigma(t)$ satisfies the $\sigma$-form of the
Painlev\'{e} III equation
\be
(t \sigma'')^2 - v_1 v_2 (\sigma')^2 + \sigma' (4 \sigma'-1) (\sigma - t \sigma') 
- \frac{1}{64} ( v_1 - v_2)^2 = 0,
\ee
with
\be
v_1 = - M+N = - \frac{2m_1}{g_s}, \qq
v_2 = M+N = - \frac{2m_2}{g_s}.
\ee
Since we are considering the $N_f=2$ case, it is natural to appear PIII${}_1$ \cite{BLMST}.
The B\"{a}cklund transformations of the PIII${}_1$ form the affine Weyl group of type $(2A_1)^{(1)}$.
The translation subgroup generates the alt-dPII equation. It generates integer shifts of parameters $v_{1,2}$. 
In terms of the gauge theory parameters, 
it corresponds to constant shifts of mass parameters $m_{1,2}$.

Let $A_n = R_n D_n$ and $B_n = R_n/D_n$. The partition function  \eqref{PartZN2} becomes
\be
\underline{Z}{}_{U(N)} = h_0^N \prod_{j=1}^{N-1} ( 1 - R_j^2 )^{N-j}. \label{PF}
\ee
Eliminating $D_n$ from \eqref{ABrec} and \eqref{ABrec2}, we obtain the recursion relation for $R_n^2$:
\bel{Rrec}
(1 - R_n^2) \Bigl( \sqrt{R_n^2 \, R_{n+1}^2 + M^2 \, \underline{g}{}_s^2 }
+ \sqrt{R_n^2\, R_{n-1}^2 + M^2 \, \underline{g}{}_s^2 } \Bigr) = 2\, n \, \underline{g}{}_s\, R_n^2.
\ee
This is equivalent to
\bel{STREQ}
\begin{split}
0=& \eta_n^2 \Bigl[ 
\xi_n^2 (1 - \xi_n)^2  - \eta_n^2 \, \xi_n^2 + \zeta^2 (1 - \xi_n)^2
\Bigr] \cr 
& +\frac{1}{2} \, \eta_n^2 \, \xi_n\, ( 1 - \xi_n)^2 (\xi_{n+1} - 2\, \xi_n + \xi_{n-1}) 
 -  \frac{1}{16} (1 - \xi_n)^4 ( \xi_{n+1} - \xi_{n-1})^2,
\end{split}
\ee
where $\xi_n \equiv R_n^2$, $\eta_n \equiv n \, \underline{g}{}_s$, $\zeta \equiv M\, \underline{g}{}_s$. 

In the planar limit $(\xi_n, \eta_n, \zeta) \rightarrow (\xi, \eta, \zeta)$, 
the second line of \eqref{STREQ} is ignored and
the three roots out of four in the resulting quartic equation in $\xi$ become
degenerate to zero at $\eta = \pm1, \zeta = 0$,
where we take the continuum limit.
In fact, setting $\xi = a^2\, u$, $\eta = \pm \,1 -(1/2)\,a^2\, t$, $\zeta =  \pm \, a^3\, M$,
we obtain at $\mathcal{O}(a^6)$
\begin{align}
	\pm t = 2u - \frac{M^2}{u^2}.
\end{align}
Eq.\eqref{STREQ} also becomes the defining relation of an algebraic variety.
With the introduction of the homogeneous coordinates $(\mathcal{X}:\mathcal{Y}:\mathcal{Z}:\mathcal{W}) = (\xi:\eta:\zeta:1)$ of 
the three-dimensional complex projective space $\mathbb{P}^3$,
this algebraic variety is the union of the hyperplane $\mathcal{Y}=0$ (with multiplicity two) and the singular K3 surface
\be
- \mathcal{Y}^2\, \mathcal{X}^2  +\mathcal{X}^2 (\mathcal{X}-\mathcal{W})^2 + (\mathcal{X}-\mathcal{W})^2 \mathcal{Z}^2= 0.
\ee
The singular loci of this surface consist of three spheres whose intersections are represented by the
$A_3$ Dynkin diagram. We are unaware of further geometrical interpretation.

In order to present the critical behavior at the planar and the double scaling limit better, let us rewrite the potential \eqref{potNf2} as
\begin{align}
	W_U (w) = \frac{N}{\widetilde{S}} \left\{ - \frac{1}{2} \left( w + \frac{1}{w} \right) + \frac{m_1 - m_2}{\Lambda_2} \log w \right\}.
\end{align}
Here, $\widetilde{S} \equiv N \underline{g}{}_s = g_s N / \Lambda_2 = - (m_1 + m_2) / \Lambda_2$ is the parameter
we fine tune to $\pm 1$,
and is the counterpart of the bare cosmological constant in $2$d gravity.
Also note that $\zeta = (m_1 - m_2) / (\widetilde{S} \Lambda_2) = \mathcal{O}(a^3)$ and the two masses are fine tuned to be equal in this limit.
It is easy to see what this critical point corresponds to in the Seiberg-Witten curve \cite{SW,HO} (quartic one), 
which is the spectral curve obtained from the planar loop equation/Virasoro constraints \cite{Dav,MM,IM}.
Omitting the standard procedure of this derivation, the curve $(y(z), z)$, where the resolvent
$\omega(z) = \displaystyle \lim_{N \rightarrow \infty} \left< g_s \sum_{I=1}^N 1/(z - w_I) \right>$ lies, is given by
\begin{align}
	y(z) &\equiv \omega(z) + \frac{{W^{(2)}}^\prime(z)}{2}, \\
	y^2 &= \frac{\Lambda_2^2}{16z^4} \left( 1 + \frac{8m_1}{\Lambda_2} z + \frac{16u}{\Lambda_2{}^2} z^2 + \frac{8m_2}{\Lambda_2} z^3 + z^4 \right).
\end{align}
Here, we have used \eqref{constNf2} and the residue relation of the resolvent at $z = \infty$.
We have parametrized the coefficient of $z^2$ by the coordinate of the moduli space of the curve.
Clearly, at our critical point $m_1 / \Lambda_2 = m_2 / \Lambda_2 = \mp 1/2$, this genus one curve shrinks to a point at $u / \Lambda_2{}^2= 3/8$.
Our limit is, therefore, the limit to the Argyres-Douglas point \cite{AD,APSW,KY}\footnote{Here,
we work in the same planar scaling limit as \cite{BLMST}}.

Let us consider the double scaling limit of \eqref{STREQ}. Let $x \equiv n/N$, 
$a^3 \equiv 1/N$ and
\be
\eta_n = \widetilde{S}\, x=1 - (1/2) a^2 \, t, \qq \zeta = a^3\, \widetilde{S}\, M, \label{DLS1}
\ee
\be
\xi(x) = \xi(n/N) =\xi_n = a^2 \, u(t).
\ee
Here, we have taken the upper sign without losing generality.
With these scaling ansatze, 
the double scaling limit is defined as the $N \rightarrow \infty$ ($a \rightarrow 0$)  limit
while simultaneously sending $\widetilde{S}$ to its critical value $1$ by \eqref{DLS1}.
The original 't Hooft expansion parameter $1/N$ gets dressed by the combination which is kept finite in this limit:
\begin{align}
	\kappa \equiv \frac{1}{N} \frac{1}{(1 - \widetilde{S})^{1 - \frac{\gamma}{2}}}, \qquad \gamma = -1
\end{align}
with $\gamma$ being the susceptibility of the system.
This last point can be checked from the free energy $F$ computation from \eqref{PF}:
\be
F = - \lim_{N \rightarrow \infty} \frac{\log \underline{Z}{}_{U(N)}}{N^2} \sim 
- \int_0^1 \de x \, (1-x) \log (1 - \xi(x)) 
\sim\left( 1 - \widetilde{S} \right)^3 = \left( 1 - \widetilde{S} \right)^{2 - \gamma}.
\ee

In the double scaling limit, the string equation \eqref{STREQ} turns into 
the Painlev\'{e} II equation
\bel{PIIu}
u'' = \frac{(u')^2}{2\, u} + u^2 - \frac{1}{2} \, t\, u - \frac{M^2}{2\, u}.
\ee
It is noteworthy that the parameter $M$ in the original model survives the limit.
We can convert \eqref{PIIu} into standard form as follows.
By using $p_u \equiv -u'/u$, this equation \eqref{PIIu} can be written as a Hamilton system with
the Hamiltonian
\bel{ham}
H_{\mathrm{II}}(u,p_u; t) = - \frac{1}{2}\, p_u^2 \, u + \frac{1}{2}\, u^2 - \frac{1}{2}\, t\, u
+ \frac{M^2}{2\, u}.
\ee
By a canonical transformation $(u,p_u) \rightarrow (v, p_v)$ with $u=-p_v$ and 
$p_u = v + (M/p_v$), this Hamiltonian becomes 
\bel{HamII}
H_{\mathrm{II}}= \frac{1}{2} \, p_v^2 + \frac{1}{2}\bigl( v^2 + t \bigr) p_v + M\, v,
\ee
and $v=v(t)$ obeys the following form of the Painlev\'{e} II equation:
\be
v'' = \frac{1}{2} \, v^3 + \frac{1}{2} \, t\, v + \left( \frac{1}{2} - M \right).
\ee
When there is no logarithmic potential ($M=0$), 
the appearance of the Painlev\'{e} II equation in the unitary matrix model
was shown in \cite{PS90a,PS90b}.

Note that \eqref{HamII} is the non-autonomous Hamiltonian
for the Painlev\'{e} II equation \cite{mal22}. The B\"{a}cklund transformations for \eqref{HamII}
are generated by \cite{oka86}
\begin{align} \label{back}
s_1(v) &= v + \frac{2\, M}{p_v},&
s_1(p_v) &= p_v,&
s_1(M) &= -M,\cr
\pi(v) &= - v,&
\pi(p_v) &= - p_v - v^2 - t,&
\pi(M) &= 1-M.
\end{align}
The restriction of $M$ to being an integer is compatible with these transformations.

We remark that these B\"{a}cklund transformations form the affine Weyl group of type $A_1^{(1)}$
and the translation $T=s_1 \pi$ generates the alternate discrete Painlev\'{e} I equation \cite{FGR93}. Explicitly, let
$v_n= T^n(v)$ and $p_n = T^n(p_v)$ ($n \in \mathbb{Z}$). Using \eqref{back},  we obtain a discrete dynamical system 
for these variables:
\be
v_{n+1} + v_n = -\frac{2(M+n)}{p_n}, \qq p_n + p_{n-1} = - v_n^2 - t.
\ee
By removing $p_n$, we find the following form of the  alt-dPI:
\be
\frac{2(M+n)}{v_n + v_{n+1}} + \frac{2(M+n-1)}{v_n + v_{n-1}} = v_n^2 + t.
\ee

%%%%%%%%%%%%%%%%  acknowledgements %%%%%%%%%%%%%%%%%%%%%%
\section*{Acknowledgments}
We thank Takayuki Koike for valuable discussions on K3 surfaces.
The work of H.~I. and T.~O. was partially supported by JSPS KAKENHI Grant Number 15K05059.

%%%%%%%%%%%%%%%%%%%%%%%%%%%%%%%%%%%%%%%%%%%%%%%%%%%%%%%%%%%

%%%%%%%%%%%%%%%%%%%%%%%%%%%%%%%%%%%%%%%%%%%%

\end{document}